\newcommand\DIMPYd{(C$_7$D$_{10}$N)$_2$CuBr$_4$}
\newcommand\DIMPY{(C$_7$H$_{10}$N)$_2$CuBr$_4$}

\documentclass[prb,twocolumn,floatfix,preprintnumbers,amsmath,amssymb]{revtex4}

\usepackage{graphicx}
\usepackage{dcolumn}
\usepackage{bm}

\begin{document}

\title{Long-lived magnons throughout the Brillouin zone of the strong-leg spin ladder \DIMPYd.}

 \author{D. Schmidiger}
 \thanks{This work is part of the Masters thesis project of David
 Schmidiger.}
 \affiliation{Neutron Scattering and Magnetism, Institute for Solid State Physics, ETH Zurich,
Switzerland.}
 \author{S. M\"uhlbauer}
 \affiliation{Neutron Scattering and Magnetism, Institute for Solid State Physics, ETH Zurich,
Switzerland.}
 \author{S. N. Gvasaliya}
 \affiliation{Neutron Scattering and Magnetism, Institute for Solid State Physics, ETH Zurich,
Switzerland.}
 \author{T. Yankova}
 \altaffiliation[Permanent address: ]{Chemistry Dept., Moscow State University, Moscow, Russia.}
 \affiliation{Neutron Scattering and Magnetism, Institute for Solid State Physics, ETH Zurich,
Switzerland.}
 \author{A. Zheludev}
 \email{zhelud@ethz.ch}
 \homepage{http://www.neutron.ethz.ch/}
 \affiliation{Neutron Scattering and Magnetism, Institute for Solid State Physics, ETH Zurich,
Switzerland.}

\begin{abstract}
Inelastic neutron scattering is used to measure spin excitations in
fully deuterated single crystal samples of the strong-leg
antiferromagnetic $S=1/2$ spin ladder compound \DIMPYd. Sharp
resolution-limited magnons are observed across the entire
one-dimensional Brillouin zone. The results validate the previously
proposed {\it symmetric} spin ladder model and provide a reliable
estimate of the relevant exchange interactions.
\end{abstract}


\maketitle

The $S=1/2$ Heisenberg antiferromagnetic (AF) spin ladder is
arguably the most important model in quantum magnetism.  Of
particular interest are spin ladders with dominant leg interactions.
They are properly described as a pair of weakly coupled spin chains.
Correspondingly, the gapped magnons are bound states of spinons
propagating on the legs.\cite{Shelton1996} This is in contrast to
strong-rung
ladders,\cite{Watson2001,Masuda2006,Fischer2011,Savici2009} where
excitations are better described in terms single-triplet transitions
in weakly interacting AF dimers on the ladder
rungs.\cite{Kolezhuk1998,Kotov1999,Normand2011} Until very recently,
the only known strong-leg spin ladder systems were
Sr$_{14}$Cu$_{24}$O$_{41}$ and its derivatives, where the magnetic
excitation spectrum was studied by inelastic
neutron\cite{Eccleston1998,Notbohm2007} and resonant X-ray
scattering\cite{Schlappa2009} techniques. Unfortunately, the energy
scales in these materials is too large to make it useful for
systematic studies of the many interesting quantum phenomena
predicted to occur in external magnetic fields or at finite
temperatures.

An almost perfect realization of a strong-leg ladder model was
recently found in the organometallic compound \DIMPY,\ DIMPY for
short.\cite{Shapiro2007,Hong2010} It is characterized by rather low
energy scales, particularly the energy gap $\Delta=0.32$~meV, and
the excellent one-dimensionality. These features enabled a direct
observation of a spin liquid to Luttinger spin liquid quantum phase
transition in applied magnetic fields in this
material.\cite{Hong2010} Inelastic neutron scattering experiments
were also performed, and the magnon dispersion relation was measured
in the vicinity of the AF zone-center. This information was then
used to estimate the ratio of Heisenberg exchange constants for the
ladder legs and rungs, respectively:
$J_\mathrm{leg}/J_\mathrm{rung}\sim 2.2$.\cite{Hong2010}
Unfortunately, the dispersion near the zone-center is not very
sensitive to this parameter and does not allow to unambiguously
establish the spin Hamiltonian. Measurements in the remaining 2/3 of
the Brillouin zone were inhibited by the use of only partially
deuterated single-crystal samples, as hydrogen scattering is a huge
contributor to neutron background. In the present study we used
fully deuterated DIMPY single crystal samples to measure the
single-magnon excitation spectrum across the entire Brilloin zone.

\begin{figure}[tbp]
\includegraphics[width=0.95\columnwidth]{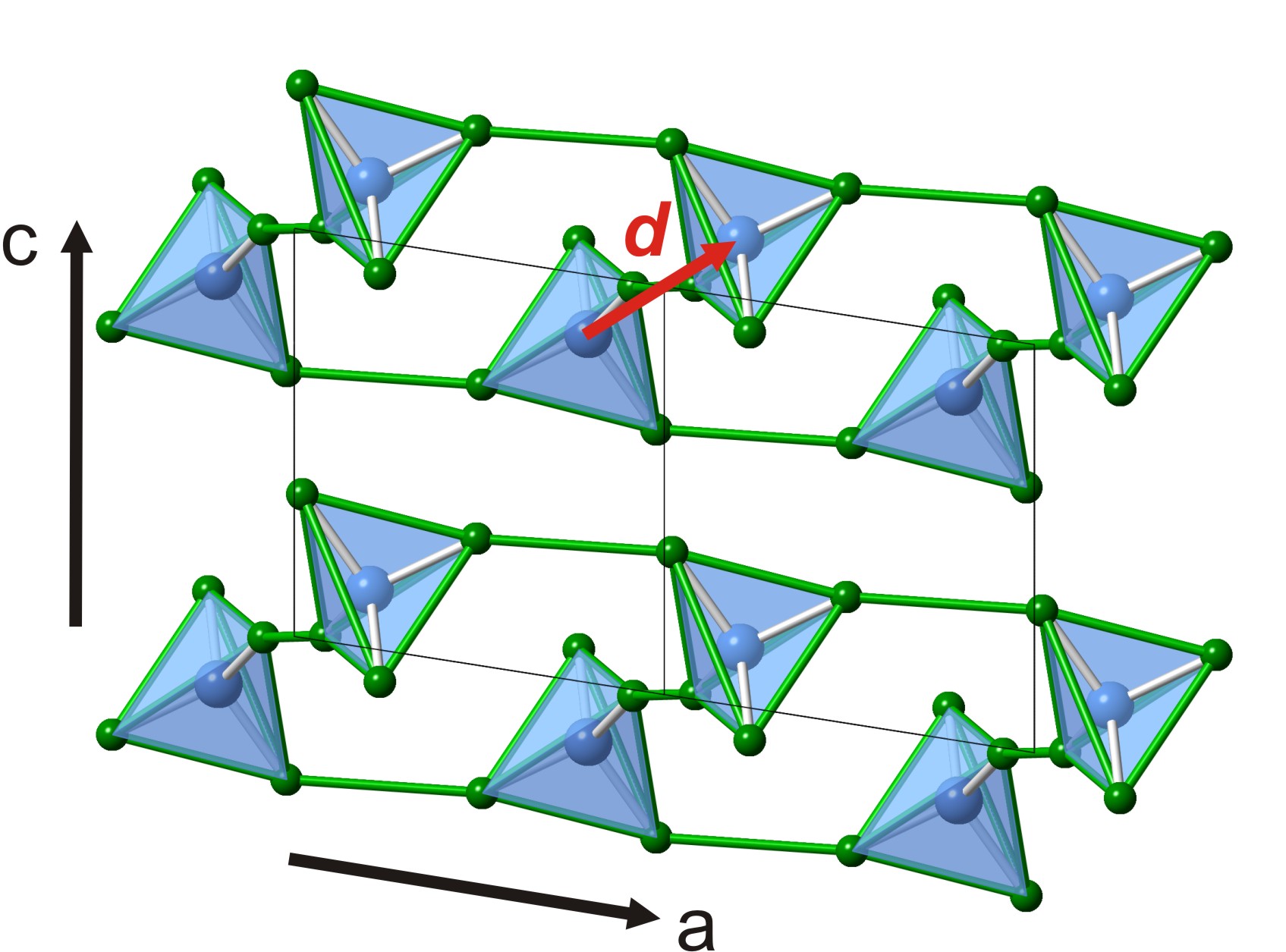}
\caption{(Color online)  A schematic representation of the
antiferromagnetic spin ladders in \DIMPYd\ in projection onto the
$(a,c)$ crystallographic plane. $S=1/2$-carrying Cu$^{2+}$ ions (
blue) are connected by double-Br$^-$ (green) bridges. The vector
$\mathbf{d}$ defines the rung of the ladder. The organic ligand
plays no role in the magnetism and is not shown.} \label{struc}
\end{figure}

\begin{figure}[tbp]
\includegraphics[width=0.95\columnwidth]{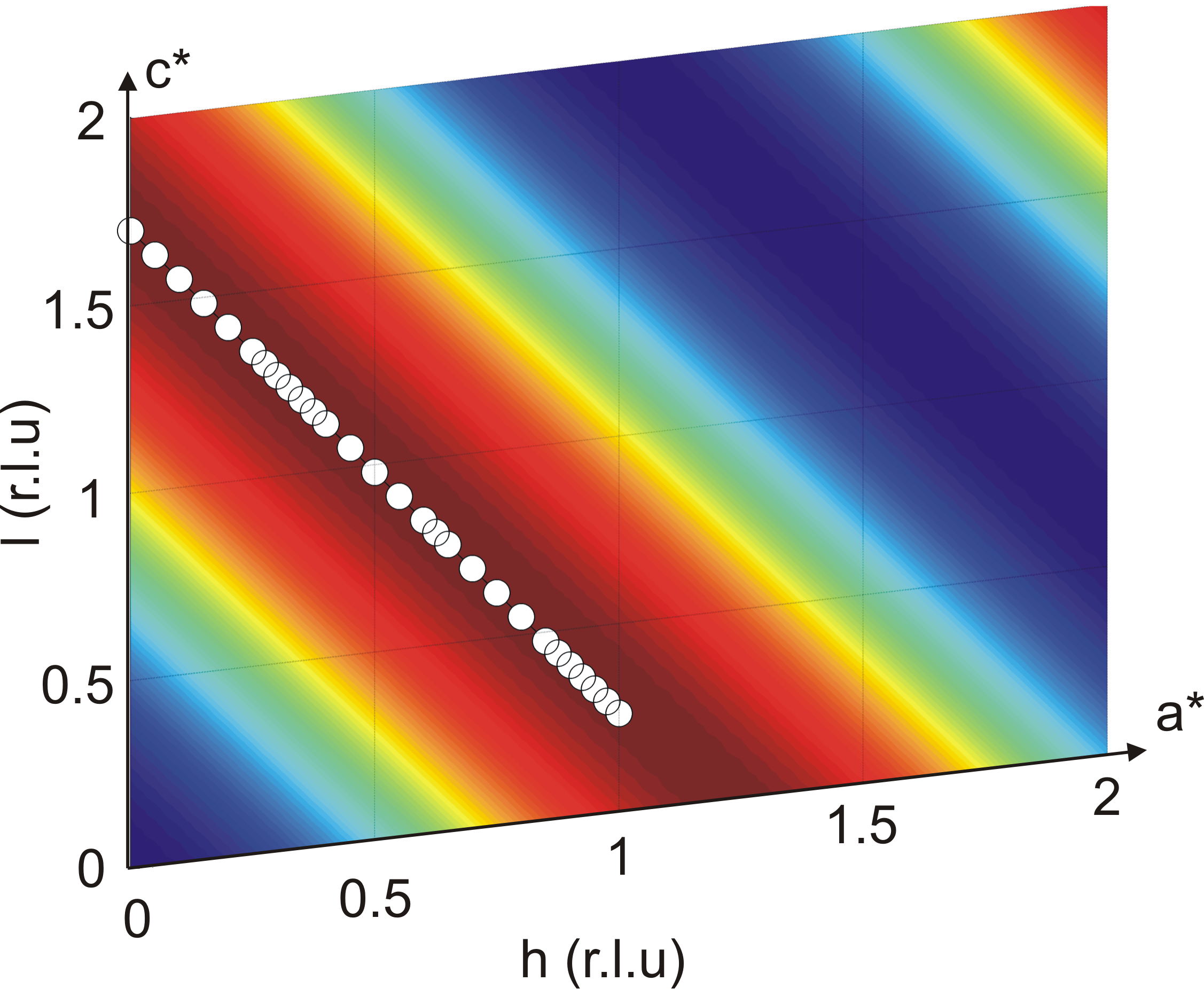}
\caption{(Color online)  False-color plot of the antiferromagnetic
rung structure factor in for \DIMPYd\, for momentum transfers in the
$(h,0,l)$ reciprocal-space plane. The color mapping ranges from zero
(dark blue), through green and yellow, to the maximum value (red).
White circles represent Symbols represent the wave vectors of all
measured inelastic neutron scans, along the $(h,0,1.7-1.44h)$ line,
as in Ref.~\protect\onlinecite{Hong2010}. The ladders are aligned
along the crystallographic $a$ axis, i.e., perpendicular to
$c^\ast$.} \label{phase}
\end{figure}

The crystal structure of DIMPY was described in
Ref.~\onlinecite{Shapiro2007}. The material is monoclinic, space
group $P2_1/n$, with room temperature lattice parameters
$a=7.50$~\AA,\ $b=31.61$~\AA,\ $c=8.20$~\AA\ and
$\beta=98.97^\circ$.\cite{Shapiro2007} As schematically shown in
Figure~\ref{struc}, the key structural feature are ladders that run
along the crystallographic $a$ direction and are composed of
magnetic Cu$^{2+}$ ions, bridged by robust Cu-Br-Br-Cu covalent
superexchange pathways. The organic ligands provide spacing between
such ladders and ensures an almost perfect magnetic one-
dimensionality. There are two types of ladders in the crystal
structure, related by symmetry. In projection onto the $(a,c)$ plane
these become equivalent. This is a crucial circumstance, since it
allows to independently measure the leg-odd and leg-even spin
excitation spectra in this compound. The total spin dynamic
structure factor for a symmetric ladder can be decomposed into its
even and odd parts:\cite{Barnes1994}
 \begin{eqnarray}
 2\mathcal{S}(\mathbf{q},\omega) & = &
 \mathcal{S}^+(\mathbf{q},\omega)
 (1+\cos\mathbf{qd})+\\ \nonumber
 & + & \mathcal{S}^-(\mathbf{q},\omega)
(1-\cos\mathbf{qd}).\label{decom}
\end{eqnarray}
Here $\mathbf{d}$ defines the ladder rung, and
$\mathcal{S}^+(\mathbf{q},\omega)$ and
$\mathcal{S}^-(\mathbf{q},\omega)$ are structure factors written for
the sum (even) and difference (odd) of spins on each rung. The
significance of this decomposition is that the single-magnon
excitations are contained in the odd channel, while the
lowest-energy leg-even excitations are a two-magnon
continuum.\cite{Barnes1994} In reciprocal space, the scattering due
to the two channels is well separated thanks to the phase factors in
Eq.~\ref{decom}. This separation has been previously made use of in
the  study of Sr$_{14}$Cu$_{24}$O$_{41}$
(Ref.~\onlinecite{Notbohm2007}) and other systems. For the
particular case of DIMPY, it is enabled for a scattering vector
$\mathbf{q}$ in the $(h,0,l)$ plane, where the product
$(\mathbf{qd})$ is the same for the two symmetry-related ladders.
The corresponding phase factor for the odd channel is plotted in
Fig.~\ref{phase}. In our experiments, following
Ref.~\onlinecite{Hong2010}, we focused on the $(h,0,1.7-1.44h)$
reciprocal-space rod, where only the leg-odd correlations are
observed and the single-magnon scattering intensity is maximized. In
the presence of any asymmetry of the spin Hamiltonian with respect
to an interchange of ladder legs, Eq.~\ref{decom} will no longer
hold. For DIMPY, simple symmetric nearest-neighbor leg and rung
exchange interactions are expected to dominate due to structural
considerations. As a result, the Hamiltonian asymmetry is
negligible. As will be discussed below, this fact is further
supported by our measurements of magnon dispersion.

\begin{figure}[tbp]
\includegraphics[width=0.95\columnwidth]{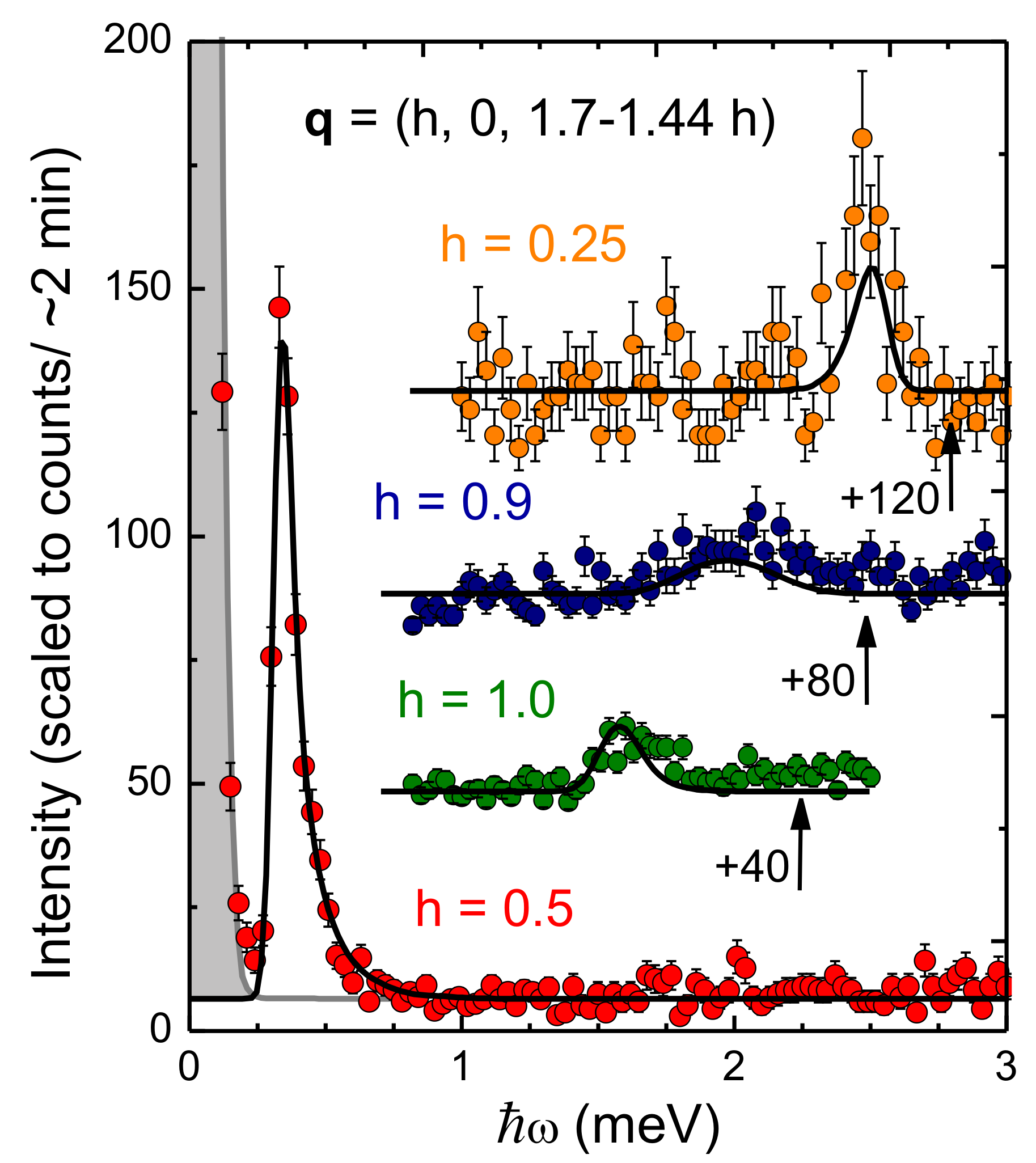}
\caption{(Color online) Symbols: typical inelastic neutron
scattering scans collected in \DIMPYd\ at the one-dimensional
antiferromagnetic zone-center, ferromagnetic zone-center and near
the antiferromagnetic zone-boundary.  The solids lines represent a
{\it simultaneous} fit of Eqs.~\protect\ref{sqw} --\protect\ref{sq}
to 28 constant-$q$ scans, as described in the text. The grayed area
is the estimated elastic incoherent background.} \label{scans}
\end{figure}

\begin{figure*}[tbp]
\includegraphics[width=0.85\textwidth]{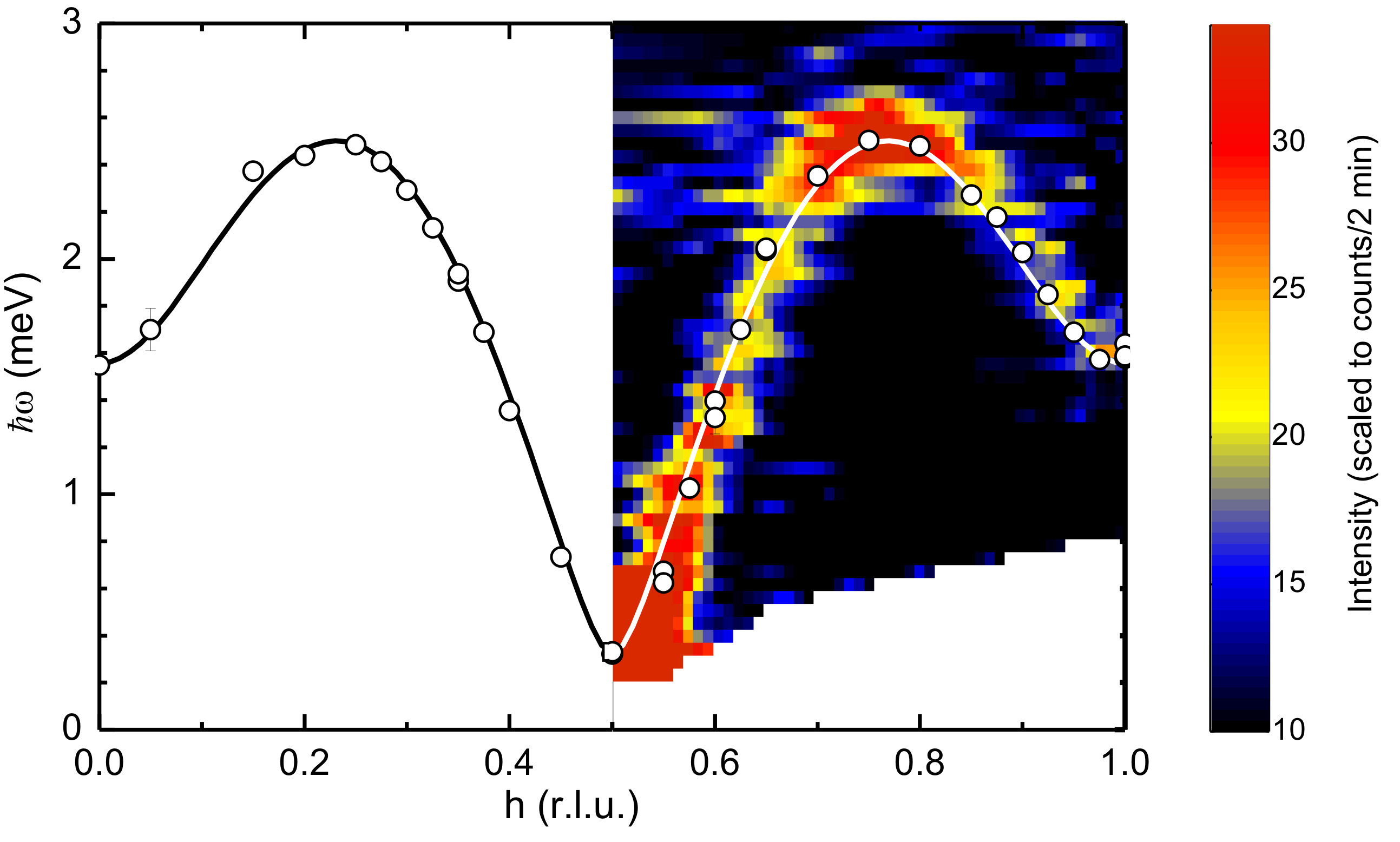}
\caption{(Color online) Dispersion of magnons measured in \DIMPYd\
for wave vectors along the  $(h,0,1.7-1.44h)$ reciprocal-space rod.
Symbols: excitation energies as determined in fits to individual
scans. Solid line: Eq.~\protect\ref{hw} with parameters obtained in
a simultaneous fit to 28 constant-$q$ scans. False-color overlay:
the bulk of experimental data at $h>0.5$.  } \label{disp}
\end{figure*}

For the present experiment, five fully deuterated single crystals
were grown from solution by slow diffusion in a temperature
gradient. The crystals, of total mass 3.7~g, were co-aligned to a
final mosaic spread of better than 1.5$^\circ$. The measurements
were performed on the TASP cold-neutron triple axis
spectrometer\cite{TASP} installed at the SINQ spallation source at
Paul Scherrer Institut and operated by ETH-Zurich. Pyrolytic
graphite monochromator and horizonatally focusing analyzer were used
in combination with a Be filter after the sample.  Most of the data
were collected in constant-$q$ scans with $E_\mathrm{f}=3.5$~meV or
$E_\mathrm{f}=4.5$~meV fixed final energy neutrons. The sample
environment was a $^3$He-$^4$He dilution refrigerator: all scans
were performed a $T=100$~mK. Typical raw scans are plotted in
symbols in Fig.~\ref{scans}. In the entire Brillouin zone the magnon
excitations are clearly visible as well-defined inelastic peaks, on
a mostly flat background of approximately 5 counts/min. The latter
was shown to be of non-magnetic origin and primarily originate from
the sample itself. The bulk of the data is shown in the false-color
plot in Fig.~\ref{disp}.

The data were analyzed using a parameterized cross section function,
written in the single mode approximation (SMA), with the empirical
dispersion relation\cite{Barnes1994} previously also used for
IPA-CuCl$_3$:\cite{Masuda2006}
\begin{eqnarray}
 \mathcal{S}(\mathbf{q},\omega)&\propto& \mathcal{S}(\mathbf{q})
 \delta(\omega-\omega_\mathbf{q}),\label{sqw}\\
 (\hbar \omega_\mathbf{q})^2&=& \Delta^2\sin^2(\pi
 h) +A^2\cos^2(\pi h)  + \nonumber\\
 &+& B^2\sin^2(2\pi h).\label{hw}
\end{eqnarray}
Here
$\mathbf{q}=h\mathbf{a}^\ast+k\mathbf{b}^\ast+l\mathbf{c}^\ast$, and
$\Delta$, $A$ and $B$ parameterize the dispersion relation,
respectively. Assuming the SMA accounts for most of the inelastic
scattering in the leg-odd channel, and that only nearest-neighbor
leg and rung interactions are relevant, the equal-time structure
factor is obtained from the Hohenberg-Brinkman first-moment sum
rule:\cite{Hohenberg74,Zaliznyak2005b}
\begin{equation}
\hbar \omega_\mathbf{q} \mathcal{S}(\mathbf{q}) \propto
-\frac{4}{3}E_\mathrm{leg}(1-\cos\mathbf{qa})-\frac{2}{3}E_\mathrm{rung}(1-\cos\mathbf{qd}).\label{sq}
\end{equation}
In this formula, $E_\mathrm{rung}=J_\mathrm{rung}\langle
\mathbf{S}_1\mathbf{S}_2\rangle$ is the mean exchange energy on the
ladder rung, and similarly for $E_\mathrm{leg}$. For all our data
$\mathbf{qd}=\pi$, so the 2nd term on the RHS is reduced to a
constant.

\begin{figure}[tbp]
\includegraphics[width=0.95\columnwidth]{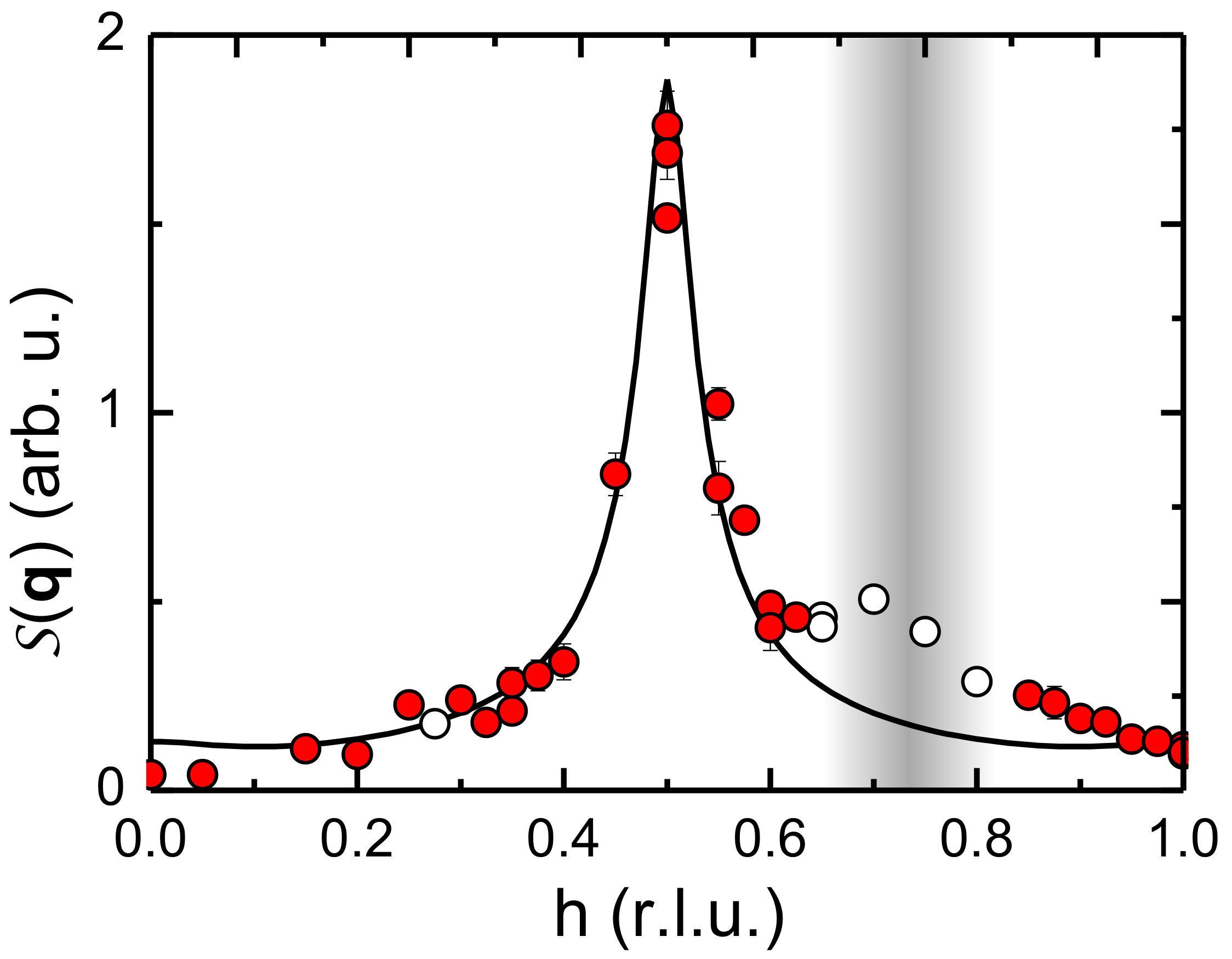}
\caption{(Color online)
 Symbols: equal-time spin structure factor $\mathcal{S}(\mathbf{q})$,
as determined in fits to individual scans. The shaded area and open
symbols show scans excluded from the global fit due to a possible
background contamination of non-magnetic origin. Solid line:
Eq.~\protect\ref{sq}, with parameters obtained in the global fit.}
\label{sqplot}
\end{figure}

The cross section was numerically convoluted\cite{Reslib} with the
spectrometer resolution function calculated in the Popovici
approximation\cite{Popovici75} and {\it globally} fit to the bulk of
available experimental data (29 constant-$q$ scans). The magnetic
form factor for Cu$^{2+}$ ions was built into the fits. In the
analysis we assumed a background consisting of a flat component and
a resolution-limited Gaussian peak at zero energy transfer, the
latter to account for elastic incoherent scattering. Scans collected
at $0.65<h<0.8$ with $E_\mathrm{f}=4.5$~meV were excluded from the
global fit due to an apparent non-dispersive feature in the
background at around 2.1~meV energy transfer at these wave vectors.
The latter was independently verified to be of non-magnetic origin
(by collecting background data at 80~K and 160~K), and shown not to
affect the data measured at equivalent wave vectors in the other
half of the Brillouin zone . The least-squares global fitting
procedure yields good agreement with $\Delta = 0.33(1)$~meV, $A =
1.55(1)$~meV, $B = 2.27(1)$~meV, and
$E_\mathrm{leg}/E_\mathrm{rung}=1.42(16)$. Scans simulated using
these parameter values are shown as solid lines in Fig.~\ref{scans}.
 The resulting dispersion relation and equal-time structure
factor are plotted in Figs.~\ref{disp} and \ref{sqplot},
respectively. Symbols in the two latter figures were obtained in
intensity- and central energy- fits to individual scans, as opposed
to global fits to all the measured data. Any remaining
discrepancies, especially in what concerns excitation intensity, may
be attributed to neutron absorbtion in the sample. This effect could
not be taken into account due to the irregular sample shape.

The measured magnon spectrum in DIMPY is markedly different from the
one previously seen in IPA-CuCl$_3$,\cite{Masuda2006,Zheludev2007}
even though the dispersion relation is similar. In the latter
material, the single-magnon branch was found to {\it terminate} at
some critical wave vector before reaching the ferromagnetic
zone-center. This effect was attributed to a two-magnon
decay.\cite{Masuda2006,Stone2006N} It is now understood that in one
dimension, this process, if only allowed by symmetry, will always
render the sharp single-magnon excitation unstable in the $\hbar
\omega-\mathbf{q}$ region spanned by two-particle momenta and
energies.\cite{Zhitomirsky2006} As was previously done for
IPA-CuCl$_3$,\cite{Masuda2006} knowing the single-particle
dispersion relation  allows us to estimate the boundaries of this
magnon-exclusion region for DIMPY:
$-\xi_\mathrm{c}+n<\mathbf{qa}/(2\pi)<\xi_\mathrm{c}+n$, where $n$
is integer and $\xi_\mathrm{c}\sim 0.85$. Experimentally, no
anomalies in the magnon branch are observed at the critical wave
vectors $n \pm \xi_\mathrm{c}$. The magnon branch persists in the
entire Brillouin zone and to within experimental error remains
resolution-limited. This implies that single-magnon states are
protected from two-magnon decay by symmetry. The relevant symmetry
operation is an interchange of the two ladder legs. Indeed, as
mentioned in the introduction, magnons are leg-odd excitations,
while two-magnon states are leg-even. Thus the stability of magnons
in DIMPY validates the symmetric-ladder model for this compound. In
contrast, in IPA-CuCl$_3$, symmetry-breaking interactions along the
ladder diagonal are known to be significant.\cite{Fischer2011}

Using the measured dispersion relation and the numerical results of
Ref.~\onlinecite{Hong2010} we can estimate
$J_\mathrm{leg}/J_\mathrm{rung}\sim 2.0$. The dominance of the leg
exchange coupling is also manifest in the experimental ratio
$E_\mathrm{leg}/E_\mathrm{rung}\sim 1.4$, independently determined
from excitation intensities. However, this latter estimate should be
treated with a degree of caution, as it relies on the the SMA. This
approximation is obviously not exact. Specifically, it does not
allow for any continuum excitations, which actually becomes
progressively more important as the relative rung strength
decreases.

In summary, our measurements confirm that DIMPY is perhaps the most
``perfect'' known spin ladder material, with dominant leg
interactions, stable magnons across the entire Brillouin zone and
experimentally accessible energy scales. The obvious next step will
be to study multi-magnon continua and the evolution of the
excitation spectrum in high magnetic fields.

\begin{acknowledgements}
This work is partially supported by the Swiss National Fund under
project 2-77060-11 and through Project 6 of MANEP. The neutron
experiments would be impossible without expert technical support
provided by Paul Scherrer Institut, particularly by Dr. Markus
Zolliker and the sample environment team. We thank Dr. V. Glazkov
(Kapitza Institute, Russian Acad. Sci.) for his involvement at the
early stages of this project.
\end{acknowledgements}


\begin{thebibliography}{21}
\expandafter\ifx\csname
natexlab\endcsname\relax\def\natexlab#1{#1}\fi
\expandafter\ifx\csname bibnamefont\endcsname\relax
  \def\bibnamefont#1{#1}\fi
\expandafter\ifx\csname bibfnamefont\endcsname\relax
  \def\bibfnamefont#1{#1}\fi
\expandafter\ifx\csname citenamefont\endcsname\relax
  \def\citenamefont#1{#1}\fi
\expandafter\ifx\csname url\endcsname\relax
  \def\url#1{\texttt{#1}}\fi
\expandafter\ifx\csname urlprefix\endcsname\relax\def\urlprefix{URL
}\fi \providecommand{\bibinfo}[2]{#2}
\providecommand{\eprint}[2][]{\url{#2}}

\bibitem[{\citenamefont{Shelton et~al.}(1996)\citenamefont{Shelton, Nersesyan,
  and Tsvelik}}]{Shelton1996}
\bibinfo{author}{\bibfnamefont{D.}~\bibnamefont{Shelton}},
  \bibinfo{author}{\bibfnamefont{A.~A.} \bibnamefont{Nersesyan}},
  \bibnamefont{and} \bibinfo{author}{\bibfnamefont{A.~M.}
  \bibnamefont{Tsvelik}}, \bibinfo{journal}{Phys. Rev. B}
  \textbf{\bibinfo{volume}{53}}, \bibinfo{pages}{8521} (\bibinfo{year}{1996}).

\bibitem[{\citenamefont{Watson et~al.}(2001)\citenamefont{Watson, Kotov,
  Meisel, Granroth, Montfrooij, Nagler, Jensen, Backov, Petruska, Fanucci
  et~al.}}]{Watson2001}
\bibinfo{author}{\bibfnamefont{B.~C.} \bibnamefont{Watson}},
  \bibinfo{author}{\bibfnamefont{V.~N.} \bibnamefont{Kotov}},
  \bibinfo{author}{\bibfnamefont{M.~W.} \bibnamefont{Meisel}},
  \bibinfo{author}{\bibfnamefont{G.~E.} \bibnamefont{Granroth}},
  \bibinfo{author}{\bibfnamefont{W.~T.} \bibnamefont{Montfrooij}},
  \bibinfo{author}{\bibfnamefont{S.~E.} \bibnamefont{Nagler}},
  \bibinfo{author}{\bibfnamefont{D.~A.} \bibnamefont{Jensen}},
  \bibinfo{author}{\bibfnamefont{R.}~\bibnamefont{Backov}},
  \bibinfo{author}{\bibfnamefont{M.~A.} \bibnamefont{Petruska}},
  \bibinfo{author}{\bibfnamefont{G.~E.} \bibnamefont{Fanucci}},
  \bibnamefont{et~al.}, \bibinfo{journal}{Phys. Rev. Lett.}
  \textbf{\bibinfo{volume}{86}}, \bibinfo{pages}{5168} (\bibinfo{year}{2001}).

\bibitem[{\citenamefont{Masuda et~al.}(2006)\citenamefont{Masuda, Zheludev,
  Manaka, Regnault, Chung, and Qiu}}]{Masuda2006}
\bibinfo{author}{\bibfnamefont{T.}~\bibnamefont{Masuda}},
  \bibinfo{author}{\bibfnamefont{A.}~\bibnamefont{Zheludev}},
  \bibinfo{author}{\bibfnamefont{H.}~\bibnamefont{Manaka}},
  \bibinfo{author}{\bibfnamefont{L.-P.} \bibnamefont{Regnault}},
  \bibinfo{author}{\bibfnamefont{J.-H.} \bibnamefont{Chung}}, \bibnamefont{and}
  \bibinfo{author}{\bibfnamefont{Y.}~\bibnamefont{Qiu}},
  \bibinfo{journal}{Phys. Rev. Lett.} \textbf{\bibinfo{volume}{96}},
  \bibinfo{pages}{047210} (\bibinfo{year}{2006}).

\bibitem[{\citenamefont{Fischer et~al.}(2011)\citenamefont{Fischer, Duffe,
  Gotz, and Uhrig}}]{Fischer2011}
\bibinfo{author}{\bibfnamefont{T.}~\bibnamefont{Fischer}},
  \bibinfo{author}{\bibfnamefont{S.}~\bibnamefont{Duffe}},
  \bibinfo{author}{\bibnamefont{Gotz}}, \bibnamefont{and}
  \bibinfo{author}{\bibfnamefont{S.}~\bibnamefont{Uhrig}},
  \bibinfo{howpublished}{arXiv:1009.3375v2} (\bibinfo{year}{2011}).

\bibitem[{\citenamefont{Savici et~al.}(2009)\citenamefont{Savici, Granroth,
  Broholm, Pajerowski, Brown, Talham, Meisel, Schmidt, Uhrig, and
  Nagler}}]{Savici2009}
\bibinfo{author}{\bibfnamefont{A.~T.} \bibnamefont{Savici}},
  \bibinfo{author}{\bibfnamefont{G.~E.} \bibnamefont{Granroth}},
  \bibinfo{author}{\bibfnamefont{C.~L.} \bibnamefont{Broholm}},
  \bibinfo{author}{\bibfnamefont{D.~M.} \bibnamefont{Pajerowski}},
  \bibinfo{author}{\bibfnamefont{C.~M.} \bibnamefont{Brown}},
  \bibinfo{author}{\bibfnamefont{D.~R.} \bibnamefont{Talham}},
  \bibinfo{author}{\bibfnamefont{M.~W.} \bibnamefont{Meisel}},
  \bibinfo{author}{\bibfnamefont{K.~P.} \bibnamefont{Schmidt}},
  \bibinfo{author}{\bibfnamefont{G.~S.} \bibnamefont{Uhrig}}, \bibnamefont{and}
  \bibinfo{author}{\bibfnamefont{S.~E.} \bibnamefont{Nagler}},
  \bibinfo{journal}{Phys. Rev. B} \textbf{\bibinfo{volume}{80}},
  \bibinfo{pages}{094411} (\bibinfo{year}{2009}).

\bibitem[{\citenamefont{Kolezhuk and Mikeska}(1998)}]{Kolezhuk1998}
\bibinfo{author}{\bibfnamefont{A.~K.} \bibnamefont{Kolezhuk}} \bibnamefont{and}
  \bibinfo{author}{\bibfnamefont{H.-J.} \bibnamefont{Mikeska}},
  \bibinfo{journal}{Int. J. Mod. Phys. B} \textbf{\bibinfo{volume}{12}},
  \bibinfo{pages}{2325} (\bibinfo{year}{1998}).

\bibitem[{\citenamefont{Kotov and Sushkov}(1999)}]{Kotov1999}
\bibinfo{author}{\bibfnamefont{V.~N.} \bibnamefont{Kotov}},
  \bibinfo{author}{\bibfnamefont{O.~P.} \bibnamefont{Sushkov}}, and R. Eder,
  \bibinfo{journal}{Phys. Rev. B} \textbf{\bibinfo{volume}{59}},
  \bibinfo{pages}{6266} (\bibinfo{year}{1999}).

\bibitem[{\citenamefont{Normand and Ruegg}(2011)}]{Normand2011}
\bibinfo{author}{\bibfnamefont{B.}~\bibnamefont{Normand}} \bibnamefont{and}
  \bibinfo{author}{\bibfnamefont{C.}~\bibnamefont{Ruegg}},
  \bibinfo{journal}{Phys. Rev. B} \textbf{\bibinfo{volume}{83}},
  \bibinfo{pages}{054415} (\bibinfo{year}{2011}).

\bibitem[{\citenamefont{Eccleston et~al.}(1998)\citenamefont{Eccleston, Uehara,
  Akimitsu, Eisaki, Motoyama, and Uchida}}]{Eccleston1998}
\bibinfo{author}{\bibfnamefont{R.~S.} \bibnamefont{Eccleston}},
  \bibinfo{author}{\bibfnamefont{M.}~\bibnamefont{Uehara}},
  \bibinfo{author}{\bibfnamefont{J.}~\bibnamefont{Akimitsu}},
  \bibinfo{author}{\bibfnamefont{H.}~\bibnamefont{Eisaki}},
  \bibinfo{author}{\bibfnamefont{N.}~\bibnamefont{Motoyama}}, \bibnamefont{and}
  \bibinfo{author}{\bibfnamefont{S.}~\bibnamefont{Uchida}},
  \bibinfo{journal}{Phys. Rev. Lett.} \textbf{\bibinfo{volume}{81}},
  \bibinfo{pages}{1702} (\bibinfo{year}{1998}).

\bibitem[{\citenamefont{Notbohm et~al.}(2007)\citenamefont{Notbohm, Ribeiro,
  Lake, Tennant, Schmidt, Uhrig, Hess, Klingeler, Behr, Buechner
  et~al.}}]{Notbohm2007}
\bibinfo{author}{\bibfnamefont{S.}~\bibnamefont{Notbohm}},
  \bibinfo{author}{\bibfnamefont{P.}~\bibnamefont{Ribeiro}},
  \bibinfo{author}{\bibfnamefont{B.}~\bibnamefont{Lake}},
  \bibinfo{author}{\bibfnamefont{D.~A.} \bibnamefont{Tennant}},
  \bibinfo{author}{\bibfnamefont{K.~P.} \bibnamefont{Schmidt}},
  \bibinfo{author}{\bibfnamefont{G.~S.} \bibnamefont{Uhrig}},
  \bibinfo{author}{\bibfnamefont{C.}~\bibnamefont{Hess}},
  \bibinfo{author}{\bibfnamefont{R.}~\bibnamefont{Klingeler}},
  \bibinfo{author}{\bibfnamefont{G.}~\bibnamefont{Behr}},
  \bibinfo{author}{\bibfnamefont{B.}~\bibnamefont{Buechner}},
  \bibnamefont{et~al.}, \bibinfo{journal}{Phys. Rev. Lett.}
  \textbf{\bibinfo{volume}{98}}, \bibinfo{pages}{027403}
  (\bibinfo{year}{2007}).

\bibitem[{\citenamefont{Schlappa et~al.}(2009)\citenamefont{Schlappa, Schmitt,
  Vernay, Strocov, Ilakovac, Thielemann, Ronnow, Vanishri, Piazzalunga, Wang
  et~al.}}]{Schlappa2009}
\bibinfo{author}{\bibfnamefont{J.}~\bibnamefont{Schlappa}},
  \bibinfo{author}{\bibfnamefont{T.}~\bibnamefont{Schmitt}},
  \bibinfo{author}{\bibfnamefont{F.}~\bibnamefont{Vernay}},
  \bibinfo{author}{\bibfnamefont{V.~N.} \bibnamefont{Strocov}},
  \bibinfo{author}{\bibfnamefont{V.}~\bibnamefont{Ilakovac}},
  \bibinfo{author}{\bibfnamefont{B.}~\bibnamefont{Thielemann}},
  \bibinfo{author}{\bibfnamefont{H.~M.} \bibnamefont{Ronnow}},
  \bibinfo{author}{\bibfnamefont{S.}~\bibnamefont{Vanishri}},
  \bibinfo{author}{\bibfnamefont{A.}~\bibnamefont{Piazzalunga}},
  \bibinfo{author}{\bibfnamefont{X.}~\bibnamefont{Wang}}, \bibnamefont{et~al.},
  \bibinfo{journal}{Phys. Rev. Lett.} \textbf{\bibinfo{volume}{103}},
  \bibinfo{pages}{047401} (\bibinfo{year}{2009}).

\bibitem[{\citenamefont{Shapiro et~al.}(2007)\citenamefont{Shapiro, Landee,
  Turnbull, Jornet, Deumal, Novoa, Robb, and Lewis}}]{Shapiro2007}
\bibinfo{author}{\bibfnamefont{A.}~\bibnamefont{Shapiro}},
  \bibinfo{author}{\bibfnamefont{C.~P.} \bibnamefont{Landee}},
  \bibinfo{author}{\bibfnamefont{M.~M.} \bibnamefont{Turnbull}},
  \bibinfo{author}{\bibfnamefont{J.}~\bibnamefont{Jornet}},
  \bibinfo{author}{\bibfnamefont{M.}~\bibnamefont{Deumal}},
  \bibinfo{author}{\bibfnamefont{J.~J.} \bibnamefont{Novoa}},
  \bibinfo{author}{\bibfnamefont{M.~A.} \bibnamefont{Robb}}, \bibnamefont{and}
  \bibinfo{author}{\bibfnamefont{W.}~\bibnamefont{Lewis}}, \bibinfo{journal}{J.
  Am. Chem. Soc.} \textbf{\bibinfo{volume}{129}}, \bibinfo{pages}{952}
  (\bibinfo{year}{2007}).

\bibitem[{\citenamefont{Hong et~al.}(2010)\citenamefont{Hong, Kim, Cotta,
  takano, Tremelling, Turnbull, landee, Kang, Christensen, Lefmann
  et~al.}}]{Hong2010}
\bibinfo{author}{\bibfnamefont{T.}~\bibnamefont{Hong}},
  \bibinfo{author}{\bibfnamefont{Y.~H.} \bibnamefont{Kim}},
  \bibinfo{author}{\bibfnamefont{C.}~\bibnamefont{Cotta}},
  \bibinfo{author}{\bibfnamefont{Y.}~\bibnamefont{takano}},
  \bibinfo{author}{\bibfnamefont{G.}~\bibnamefont{Tremelling}},
  \bibinfo{author}{\bibfnamefont{M.~M.} \bibnamefont{Turnbull}},
  \bibinfo{author}{\bibfnamefont{C.~P.} \bibnamefont{landee}},
  \bibinfo{author}{\bibfnamefont{H.-J.} \bibnamefont{Kang}},
  \bibinfo{author}{\bibfnamefont{N.~B.} \bibnamefont{Christensen}},
  \bibinfo{author}{\bibfnamefont{K.}~\bibnamefont{Lefmann}},
  \bibnamefont{et~al.}, \bibinfo{journal}{Phys. Rev. Lett.}
  \textbf{\bibinfo{volume}{105}}, \bibinfo{pages}{137207}
  (\bibinfo{year}{2010}).

\bibitem[{\citenamefont{Barnes and Riera}(1994)}]{Barnes1994}
\bibinfo{author}{\bibfnamefont{T.}~\bibnamefont{Barnes}} \bibnamefont{and}
  \bibinfo{author}{\bibfnamefont{J.}~\bibnamefont{Riera}},
  \bibinfo{journal}{Phys. Rev. B} \textbf{\bibinfo{volume}{50}},
  \bibinfo{pages}{6817} (\bibinfo{year}{1994}).

\bibitem{TASP} F. Semadeni, B. Roessli, and P. Böni, Physica B {\bf 297}, 152
(2001).


\bibitem[{\citenamefont{Hohenberg and Brinkman}(74)}]{Hohenberg74}
\bibinfo{author}{\bibfnamefont{P.~C.} \bibnamefont{Hohenberg}}
  \bibnamefont{and} \bibinfo{author}{\bibfnamefont{W.~F.}
  \bibnamefont{Brinkman}}, \bibinfo{journal}{Phys. Rev. B}
  \textbf{\bibinfo{volume}{10}}, \bibinfo{pages}{128} (\bibinfo{year}{1974}).

\bibitem[{\citenamefont{Zaliznyak and Lee}(2005)}]{Zaliznyak2005b}
\bibinfo{author}{\bibfnamefont{I.}~\bibnamefont{Zaliznyak}} \bibnamefont{and}
  \bibinfo{author}{\bibfnamefont{S.-H.} \bibnamefont{Lee}},
  \emph{\bibinfo{title}{Modern Techniques of Characterizing Magnetic
  Materials}} (\bibinfo{publisher}{Kluwer Academic}, \bibinfo{year}{2005}),
  chap. \bibinfo{chapter}{1.3.6.2}, p.~\bibinfo{pages}{41}.

\bibitem[{\citenamefont{Zheludev}(2009)}]{Reslib}
\bibinfo{author}{\bibfnamefont{A.}~\bibnamefont{Zheludev}},
  \emph{\bibinfo{title}{Reslib resolution library for matlab}},
  \bibinfo{howpublished}{\texttt{http://www.neutron.ethz.ch/research/resources/reslib}}
  (\bibinfo{year}{2009}).

\bibitem[{\citenamefont{Popovici}(1975)}]{Popovici75}
\bibinfo{author}{\bibfnamefont{M.}~\bibnamefont{Popovici}},
  \bibinfo{journal}{Acta Cryst.} \textbf{\bibinfo{volume}{A31}},
  \bibinfo{pages}{507} (\bibinfo{year}{1975}).

\bibitem[{\citenamefont{Zheludev et~al.}(2007)\citenamefont{Zheludev, Garlea,
  Masuda, Manaka, Regnault, Ressouche, Grenier, Chung, Qiu, Habicht
  et~al.}}]{Zheludev2007}
\bibinfo{author}{\bibfnamefont{A.}~\bibnamefont{Zheludev}},
  \bibinfo{author}{\bibfnamefont{V.~O.} \bibnamefont{Garlea}},
  \bibinfo{author}{\bibfnamefont{T.}~\bibnamefont{Masuda}},
  \bibinfo{author}{\bibfnamefont{H.}~\bibnamefont{Manaka}},
  \bibinfo{author}{\bibfnamefont{L.-P.} \bibnamefont{Regnault}},
  \bibinfo{author}{\bibfnamefont{E.}~\bibnamefont{Ressouche}},
  \bibinfo{author}{\bibfnamefont{B.}~\bibnamefont{Grenier}},
  \bibinfo{author}{\bibfnamefont{J.-H.} \bibnamefont{Chung}},
  \bibinfo{author}{\bibfnamefont{Y.}~\bibnamefont{Qiu}},
  \bibinfo{author}{\bibfnamefont{K.}~\bibnamefont{Habicht}},
  \bibnamefont{et~al.}, \bibinfo{journal}{Phys. Rev. B}
  \textbf{\bibinfo{volume}{76}}, \bibinfo{pages}{054450}
  (\bibinfo{year}{2007}).

\bibitem[{\citenamefont{Stone et~al.}(2006)\citenamefont{Stone, Zaliznyak,
  Hong, Broholm, and Reich}}]{Stone2006N}
\bibinfo{author}{\bibfnamefont{M.~B.} \bibnamefont{Stone}},
  \bibinfo{author}{\bibfnamefont{I.~A.} \bibnamefont{Zaliznyak}},
  \bibinfo{author}{\bibfnamefont{T.}~\bibnamefont{Hong}},
  \bibinfo{author}{\bibfnamefont{C.~L.} \bibnamefont{Broholm}},
  \bibnamefont{and} \bibinfo{author}{\bibfnamefont{D.~H.} \bibnamefont{Reich}},
  \bibinfo{journal}{Nature} \textbf{\bibinfo{volume}{440}},
  \bibinfo{pages}{190} (\bibinfo{year}{2006}).

\bibitem[{\citenamefont{Zhitomirsky}(2006)}]{Zhitomirsky2006}
\bibinfo{author}{\bibfnamefont{M.}~\bibnamefont{Zhitomirsky}},
  \bibinfo{journal}{Phys. Rev. B} \textbf{\bibinfo{volume}{73}},
  \bibinfo{pages}{100404} (\bibinfo{year}{2006}).

\end{thebibliography}

\end{document}